\documentclass[preprint,authoryear,12pt]{elsarticle}

\usepackage{amssymb}
\usepackage{amsmath}

\journal{Ecological Complexity}

\begin{document}

\begin{frontmatter}



\title{Mean-field theory of collective motion due to velocity alignment}
\author{Pawel Romanczuk\fnref{newaddress}}
\author{Lutz Schimansky-Geier}
\fntext[newaddress]{Present address: Max Planck Institute for the Physics of Complex Systems, N{\"o}thnitzerstr 38, 01187 Dresden, Germany}
\address{Department of Physics, Humboldt Universit{\"a}t zu Berlin, Newtonstr. 15, 12489 Berlin, Germany}
\begin{abstract}
Establishing a direct link between individual based models and the corresponding population description is a common challenge in theoretical ecology. Swarming is a prominent example, where collective effects arising from interactions of individuals are essential for the understanding of large-scale spatial population dynamics, and where both levels of modelling have been often employed without establishing this connection. 

Here, we consider a system of self-propelled agents with velocity alignment in 2D and derive a mean-field theory from the microscopic dynamics via a nonlinear Fokker-Planck equation and a moment expansion of the probability density. 
We analyze the stationary solutions corresponding to macroscopic collective motion (ordered state) and the disordered solution with no collective motion in the spatially homogeneous system. In particular, we discuss the impact of different propulsion functions governing individual dynamics. 
Our results predict a strong impact of individual dynamics on the mean field onset of collective motion (continuous vs discontinuous). In addition to the macroscopic density and velocity fields, we consider the effective ``temperature'' field, measuring velocity fluctuations around the mean velocity. We show that the temperature decreases strongly with increasing level of collective motion despite constant fluctuations on individual level, which suggests that extreme caution should be taken in deducing individual behavior, such as, state-dependent individual fluctuations from mean-field measurements [Yates {\em et al.} (2009) Proc. Natl. Acad. Sci. USA 106:5464-5469].   

\end{abstract}


\end{frontmatter}


\section{Introduction}

Collective motion of living organisms, such as exhibited by bird flocks, fish schools or insect swarms is an ubiquitous and fascinating self-
organization phenomenon, which has attracted scientists with very different backgrounds, ranging from biology and ecology to mathematics, physics and engineering. 

The understanding of grouping and collective behavior is essential for our understanding of large scale dynamics of populations \citep{okubo_diffusion_2001,krause_living_2002}. The effective interaction of individuals leading to the 
onset of collective motion of up to thousands or millions of individuals, 
does not only have a strong impact on dispersal and 
migration of entire populations, but poses also interesting theoretical questions 
about modelling and the corresponding mathematical description of such ecological systems \citep{parrish_complexity_1999}. 
The individual based -- or ``Langragian'' -- approach provides a natural 
framework for a detailed description of individual dynamics in terms of (stochastic) equations of motion. The downside of the approach is the difficulty to obtain analytical results on the behavior of many interacting individuals. Thus, the approach often relies on extensive numerical simulations, which make it difficult to gain a profound understanding of the general dynamical behavior of the system. Furthermore, despite the continuous progress in the development and application of novel computational techniques, large scale simulation of individual based model on ecological length and time scales are still extremely expensive in terms of computational time. 
Thus, at the level of populations, it seems more reasonable to take the ``Eulerian'' viewpoint, where the behavior of the ecological system is described by partial differential equations governing the dynamics of mean field observables such as the population density, which may even allow us to obtain analytical results. 

It is obvious that the different viewpoints are not independent: populations consist of individuals; their dynamics origin from individual behavior. The question is, whether, for a given microscopic individual based model, a corresponding macroscopic or mean-field description can be derived. Depending on the complexity of the individual based model this might be not feasible, but for simple models which focus on few essential mechanisms, it is possible to establish a direct link between the two levels of description.

Here, starting from a simple individual based model of collective motion, we will derive systematically the corresponding mean-field equations. We consider a model of individuals interacting only via a velocity-alignment interaction, which tends to harmonize the speeds and directions of motion of neighboring individuals. The description of individuals dynamics is based on the concept of so-called Active Brownian particles, which can be described by stochastic equations of motion with a velocity-dependent friction function which,  depending on the velocity, may also assume negative values \citep{schweitzer_complex_1998,erdmann_brownian_2000,schweitzer_brownian_2003}. Thus, in the following we will refer to individuals as ``particles'' or ``agents''. 

Two types of velocity alignment can be distinguished: Particles with nematic interaction align either parallel or anti parallel, whereas a polar interaction acts strictly towards parallel alignment of individual velocity vectors (see \citealp{ramaswamy_mechanics_2010} for a review). The velocity-alignment interaction studied here is polar \citep{czirok_formation_1996} and can be seen as a continuous version of the well known Vicsek-model \citep{vicsek_novel_1995,chate_collective_2008}. It reduces for pairwise interaction of self-propelled particles with constant speed to the polar-alignment model studied by \citet{peruani_mean-field_2008}. 
Recently, there has been a number of publication on the coarse-grained description of self-propelled particles with velocity alignment. The ansatz proposed by \citet{toner_flocks_1998,toner_long-range_1995} is based on the formulation of mesoscopic equations of motion for the density and velocity fields using symmetry and conservation laws  (see also, e.g., \citealp{toner_hydrodynamics_2005,ramaswamy_mechanics_2010}). Recently, \citet{bertin_boltzmann_2006,bertin_microscopic_2009} derived hydrodynamic equations of interacting self-propelled particles by a Boltzmann approach. An alternative derivation of kinetic equations for the Vicsek model, was presented by \citet{ihle_kinetic_2011}. Ihle  derives the corresponding transport coefficients by taking into account multiparticle collisions and shows that the corresponding results are in good agreement with numerical simulations of the Vicsek model. Bertin {\em et al.} as well as Ihle assume in their derivations of  coarse-grained description to be close to the critical point for the onset of collective motion (small center of mass velocity).  

Here, in contrast to the Vicsek model, we consider individual dynamics in terms of stochastic differential equation (Langevin equation) with continuous time. We do not assume a constant speed or restrict to the vicinity of the critical point for the onset of collective motion. 
The mean field equations are derived in a systematic way starting from the microscopic Langevin equations by formulating the corresponding Fokker-Planck equation for the probability density. Hereby, we consider directly multi-individual interactions as each individual couples to the mean velocity of its neighbours. In addition to the density and velocity fields, we consider explicitly the effective temperature field of the active Brownian particle gas.
In this work, we focus on the impact of different (nonlinear) velocity-dependent friction functions on the onset of collective motion. 

Although the kinetic theory derived in this paper, can in principle be used to analyze the spatially inhomogeneous solutions, we restrict ourselves here to the discussion  of the spatially homogeneous case. 
Please note that recent publications show the instability of the homogeneous solution in large systems with short-ranged velocity alignment \citep{bertin_boltzmann_2006,bertin_microscopic_2009,lee_fluctuation-induced_2010,ihle_kinetic_2011}. Thus, the results presented here can not be generalized to the so-called thermodynamic limit with local interactions. 

The derivation of the kinetic theory is based on the formulation of moment equations of the corresponding probability distribution. This approach has been employed, for example, by \citet{riethmueller_langevin_1997} to analyze the behavior of a quasi one-dimensional granular system. \citet{erdmann_kollektive_2003} used it to analyze the mean field of non-interacting active particles with Rayleigh-Helmholtz friction. Only recently, \citet{romanczuk_collective_2010} used the approach to analyze the collective motion of active particles in one spatial dimension. 

In general, for a system far from equilibrium the probability distribution is not Gaussian and a correct description requires infinitely many moments (see for example \citealp{pawula_approximation_1967,pawula_approximating_1987}). Thus, depending on the detailed model, it may be necessary to neglect higher moments in order to obtain a closure of the system of moment equations. The approximation of a non-Gaussian probability distribution by a finite number of moments may lead to unphysical behavior, such as negative values or artificial oscillations of the (approximated) probability distribution. Therefore, we will compare our analytical results to numerical simulations of the microscopic system. 

\section{Active Brownian particles with velocity alignment}
We consider a system of $N$ individuals with mass $m$ in two spatial dimensions. For simplicity, we neglect the finite size of individuals and consider them to be point-like active Brownian particles. The evolution of the particle positions ${\bf r}_i={\bf r}_i(t)$ and velocities ${\bf v}_i={\bf v}_i(t)$ is described by the following set of stochastic equations of motion ($i=1\dots N$):
\begin{align}
\frac{{\text{d}\bf r}_i}{\text{d} t} & = {\bf v}_i \label{eq:va_r}\\
m \frac{{\text{d}\bf v}_i}{\text{d} t} & = -\gamma({\bf v}_i){\bf v}_i + \mu ({\bf u}_{\varepsilon,i} - {\bf v}_i) +\sqrt{2 D} {\boldsymbol \xi}_i(t) \label{eq:va_v}
\end{align}
The first term on the right hand side of Eq. \eqref{eq:va_v} is a friction/propulsion force, which describes the deterministic velocity dynamics of non-interacting active particles. This velocity-dependent friction is negative at low speeds, which leads to an acceleration of the individual. As a consequence an individual has a preferred speed different from zero. In the following sections, we will discuss active particles with two different velocity-dependent friction functions: the nonlinear Rayleigh-Helmholtz friction \citep{rayleigh_theory_1894,erdmann_brownian_2000}, as well as a variant of the linear Schienbein-Gruler friction \citep{schienbein_langevin_1993,erdmann_brownian_2000}.

\begin{figure}
\begin{center}
  \includegraphics[width=0.6\linewidth]{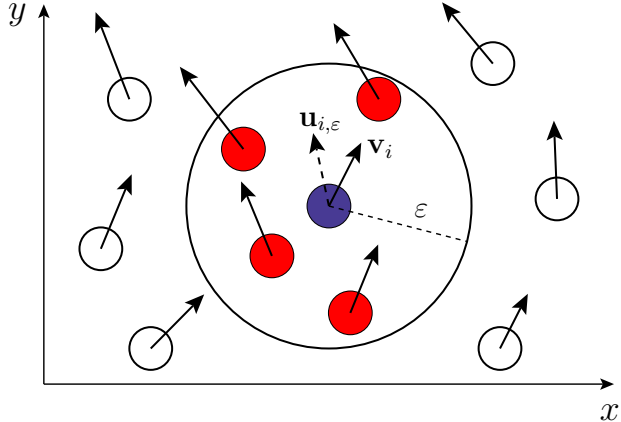} 
\end{center}
\caption{Scheme of the velocity-alignment interaction. The focal individual (blue/dark grey) interacts with all neighbors (red/bright grey) within the metric interaction range $\varepsilon$.\label{fig:scheme_va}}
\end{figure}

The second term is a velocity-alignment interaction \citep{niwa_self-organizing_1994,okubo_diffusion_2001,romanczuk_collective_2010}, with 
\begin{align}
{\bf u}_{\varepsilon,i}={\bf u}_{\varepsilon,i}({\bf r_i},t)=\frac{1}{N_{\varepsilon,i}} \sum_{j=1}^{N_{\varepsilon,i}} {\bf v}_j
\end{align}  
being the mean velocity of particles within the neighborhood $S_{\varepsilon,i}$ defined via the radius $\varepsilon>0$ around the focal particle\footnote{Please note that the sum includes the velocity of the focal particle ${\bf v}_i$. Thus, for a solitary particle $N_\varepsilon=1$ and ${\bf u}_{\varepsilon,i}={\bf v}_i$.}: ${\bf r_j}\in S_{\varepsilon,i}$ if $|{\bf r}_i-{\bf r}_j|<\varepsilon$. The alignment acts towards harmonization of the velocity of the focal particle with the local average velocity of its neighbors. The alignment strength $\mu=1/\tau_a>0$ determines the relaxation time $\tau_a$ of the velocity of the focal particle towards the average velocity of surrounding particles (see Fig. \ref{fig:scheme_va}). For solitary particles, with no neighbors, or a system in a perfectly ordered state where all particles move with exactly the same velocity, the alignment force vanishes as ${\bf u}_{\varepsilon,i}={\bf v}_i$. On the other hand, for a large number of neighbors $N_{\varepsilon,i}\gg 1$ moving with random velocities (disordered state), the mean velocity vanishes ${\bf u}_{\varepsilon,i}={\bf 0}$ and the velocity alignment force leads to an additional ``social'' friction $-\mu {\bf v}_i$.

The last term on the right hand side of Eq. \eqref{eq:va_v} accounts for is stochastic force, which introduces a random component into the motion of each individual. For each individual it is given by a Gaussian random force with intensity $D$ independent on the current velocity of the focal individual as well as on the motion of other individuals.  Here, ${\boldsymbol \xi}(t)$ is a Cartesian vector with uncorrelated components given by a stochastic Gaussian process: $\langle \xi_k(t)\rangle =0$, $\langle \xi_l(t)\xi_k(t')\rangle=\delta_{kl}\delta(t-t')$.

Throughout this work we will consider the  motion of individuals in a rectangular spatial domain of the size $L^2$ with periodic boundary condition (torus). Furthermore we rescale all forces by the mass of individuals, by setting $m=1$.

We introduce the probability distribution $P({\bf r},{\bf v},t)\,d{\bf r}\,d{\bf v}$, which determines the probability to find an individual at time $t$, in the spatial region $[{\bf r},{\bf r}+d{\bf r}]$ moving with velocity within the interval $[{\bf v},{\bf v}+d{\bf v}]$. 
In the following, we make the mean field assumption for the alignment term $N_{\varepsilon,i}\gg 1$ and derive the mean field theory corresponding to the microscopic dynamics given in Eqs. \eqref{eq:va_r}, \eqref{eq:va_v} via the formulation of moment equations of the corresponding probability distribution. 
We define the $n$-th moment of the $k$-component of the velocity vector $ v_k = {\bf v}{\bf e}_k$, with ${\bf e}_k$ being a canonical basis unit vector as
\begin{align}
\langle v^n_k \rangle({\bf r},t) = \frac{1}{\rho({\bf r},t)}\int d{\bf v} v_k^n P({\bf r},{\bf v},t)\,,
\end{align}
with $k=x,y$ and $n=0,1,2,\dots$. The spatial density of individuals $\rho({\bf r},{\bf v},t)$ corresponds to the zeroth moment ($n=0$) and gives us the normalization of the probability density with respect to integration over the velocity space:
\begin{align}
\rho({\bf r},t)=\int d{\bf v} P({\bf r},{\bf v},t)\,.
\end{align}
Multiplying the $n$-th moment by the density and taking the derivative with respect to time, we obtain the dynamics of the velocity moments:
\begin{equation}
\frac{\partial}{\partial t}(\rho \langle v_k^n \rangle) = \int d{\bf v} \, v_k^n \frac{\partial P}{\partial t}.
\label{eq:n-moment}
\end{equation}
For more than one dimension the above definition can easily be extended to mixed moments as, for example, the covariance:
\begin{equation}
\langle v_x^n v_y^m\rangle = \frac{1}{\rho} \int d{\bf v} \, v_x^n v_y^m P({\bf r},{\bf v},t) \quad n,m > 0.
\label{eq:mixedmoments}
\end{equation}

The Fokker-Planck equation, which determines the evolution of the distribution function of individuals $P$, for the microscopic dynamics given in Eq. \eqref{eq:va_v} reads
\begin{align}\label{eq:fpeRH2d}
\frac{\partial}{\partial t} P({\bf r},{\bf v},t) =   -{\bf v}\nabla_{\bf r} P -\nabla_{\bf v}\left\{\left[-\gamma({\bf v}){\bf v}+\mu \left({\bf u}_{\varepsilon}- {\bf v}\right)\right]P\right\} 
 + D \Delta_{\bf v} P  \,.
\end{align}
The local average velocity ${\bf u}_{\varepsilon}={\bf u}_{\varepsilon}({\bf r},{\bf v},t)$ in the alignment term depends implicitly on the probability density $P$, thus the above equation is a nonlinear Fokker-Planck equation as discussed, e.g., by \citet{frank_nonlinear_2005}.
In the continuous description we may express ${\bf u}_{\varepsilon}$ as an integral over the distribution function:
\begin{align}\label{eq:mf_va_vel}
{\bf u}_{\varepsilon}({\bf r},t)=\frac{1}{\int_{S_\varepsilon({\bf r})} \rho({\bf r}',t) d{{\bf r}'}}\int_{S_\varepsilon({\bf r})} d{\bf r}' \int d{\bf v}' {\bf v}' P({\bf r}',{\bf v}',t)   \,.
\end{align}
$S_\varepsilon$ represents the spatial neighbourhood of the position ${\bf r}$, defined via a metric distance: ${\bf r'}\in S_{\varepsilon}$ if $|{\bf r}-{\bf r'}|<\varepsilon$. 

Please note that in the limit $\varepsilon\to0$, the local average velocity in the alignment term ${\bf u}_\varepsilon$ reduces directly to the mean field velocity ${\bf u}({\bf r},t)=(\langle v_x \rangle, \langle v_y \rangle)$, whereas for a finite value of $\varepsilon$ it can be seen as an approximation of ${\bf u}$ under the assumption of a constant density on the length-scale corresponding to $\varepsilon$.

\section{Rayleigh-Helmholtz friction function}
At first we derive the mean field theory for active Brownian particles with the so called Rayleigh-Helmholtz friction, which reads
\begin{align}\label{eq:RHfriction}
-\gamma({\bf v}){\bf v}= (\alpha - \beta {\bf v}^2){\bf v},
\end{align}
with $\alpha,\beta\geq0$. The above friction (or propulsion) function in the has been studied in detail in the context of active Brownian motion \citep{erdmann_brownian_2000,erdmann_excitation_2002,schimansky-geier_stationary_2005}. It was used in general models of swarm dynamics \citep{mikhailov_noise-induced_1999,rappel_self-organized_1999,erdmann_noise_2005,dorsogna_self-propelled_2006,ebeling_swarm_2008} as well as for modeling of fish schools \citep{niwa_self-organizing_1994,niwa_newtonian_1996}, bacterial swarming \citep{romanczuk_beyond_2008} and locusts swarms \citep{bazazi_nutritional_2010}. The term $\alpha {\bf v}$ in \eqref{eq:RHfriction} leads to a acceleration of individuals at low velocities in the direction of motion, whereas the term $-\beta {\bf v}^2{\bf v}$ represents a nonlinear-friction. The deterministic, stationary speed of individuals can be calculated from the balance of the acceleration and friction to $v_0=\sqrt{\alpha/\beta}$.    

In analogy to the one-dimensional case \citep{romanczuk_collective_2010} we insert Eq. \eqref{eq:fpeRH2d} with Eq. \eqref{eq:RHfriction} into Eq. \eqref{eq:n-moment} and obtain for the $n$-th moment of $v_x$
\begin{align}
\frac{\partial}{\partial t}(\rho \,\langle v^n_x\rangle) & =  
\int v_x^n  \Biggl\{  - v_x \frac{\partial}{\partial x} - v_y \frac{\partial}{\partial y} \nonumber \\ 
			& \qquad\qquad - \frac{\partial}{\partial v_x} \left[-\left(\alpha - \beta (v_x^2+v_y^2)\right) v_x +\mu(u_{\varepsilon,x}-v_x)\right] \nonumber \\
			& \qquad\qquad - \frac{\partial}{\partial v_y} \left[-\left(\alpha - \beta (v_x^2+v_y^2)\right) v_y +\mu(u_{\varepsilon,y}-v_y)\right] \nonumber \\ 
 			& \qquad\qquad + D \left[ \frac{\partial^2}{\partial v_x^2}+\frac{\partial^2}{\partial v_y^2}  \right] \Biggr\} P d{\bf v}.
\end{align}
Since the probability distribution approaches zero at infinity $\lim_{v_k\to \pm \infty}P=0$, the terms with derivatives with respect to $v_y$ vanish. The terms with derivatives with respect to $v_x$ can be partially integrated and we obtain
\begin{align}\label{eq:RHva2d_mom}
\frac{\partial}{\partial t}(\rho\, \langle v^n\rangle) = & 
			- \frac{\partial}{\partial x} \left(\rho \ \langle v_x^{n+1} \rangle\right) - \frac{\partial}{\partial y} \left(\rho \ \langle v_x^n \ v_y \rangle\right)  \nonumber \\
 		& + n \ \alpha \ \rho \ \langle v_x^n \rangle -  \ n \ \beta \ \rho \ \left(\langle v_x^{n+2} \rangle + \langle v_x^n \ v_y^2 \rangle \right) \nonumber \\
 &  + n \ \left(n-1\right) \ D \ \rho \ \langle v_x^{n-2} \rangle.
\end{align}

Please note that the dynamics of the $n$-th moment depend on the $n+1$-th moment. Thus, we obtain an infinite hierarchy of coupled moment equations, for which we have to define a closure condition. 

We rewrite the velocity vector as a sum: ${\bf v}={\bf u}+\delta{\bf v}$, with a mean-field velocity ${\bf u}=(u_x,u_y)$ and a vector of symmetric deviations around the mean $\delta {\bf v}=(\delta v_x,\delta v_y)$ with $\langle \delta v_k^a \rangle=0$ for odd $a$. . Furthermore, we assume that the deviations in $x$ and $y$ are independent, thus
\begin{align}
\langle \delta v_x^n \delta v_y^m \rangle = \langle \delta v_x^n \rangle\langle \delta v_y^m \rangle.
\end{align}
We have verified this assumption by numerical simulations, which show negligible mean-field cross-correlations of velocity deviations. This can be also justified by symmetry considerations of the two possible stationary states discussed in detail further below: the completely isotropic disordered state with vanishing mean velocity $|{\bf u}|=0$, and the ordered state with $|{\bf u}|>0$, where the cross-correlations of the deviations parallel and perpendicular to the mean field velocity must vanish. 

From the above assumption we obtain a diagonal covariance matrix:
\begin{align}
{{\bf M}_\text{cv}} & = \left( \begin{array}{cc} \langle \delta v_x^2 \rangle & \langle \delta v_x \delta v_y \rangle \\ \langle \delta v_y \delta v_x \rangle & \langle \delta v_y^2 \rangle \end{array}\right) = \left( \begin{array}{cc} T_x & 0 \\ 0 & T_y \end{array}\right). 
\end{align}
In analogy to kinetic gas theory we will refer to the ``vector'' of the non-vanishing diagonal elements ${\bf T}=(T_x,T_y)$ as temperature. Please note that, in contrast to simple gas at thermal equilibrium the components of ${\bf T}$ may differ at non-equilibrium steady state: $T_x\neq T_y$.

Using ${\bf u}$ and ${\bf T}$ we may write the moments $\langle v_k^n\rangle$ as:
\begin{subequations}\label{eq:simplemoments}
\begin{align}
\langle v_k \rangle   & = u_k \\
\langle v_k^2 \rangle & = u_k^2 + T_k \\
\langle v_k^3 \rangle & = u_k^3 + 3 \ u_k \ T_x \\
\langle v_k^4 \rangle & = u_k^4 + 6 \ u_k^2 \ T_k + T_k^2 + \theta_k, 
\end{align}
\end{subequations}
with $k=x,y$ and $\theta_k$ being the mean field temperature fluctuations in $k$-direction  \citep{erdmann_kollektive_2003,romanczuk_collective_2010} defined as:
\begin{align}\label{eq:theta}
\theta_k = \langle \left(\left(v_k - u_k\right)^2 - T_k\right)^2 \rangle= \langle \delta v_k^4 \rangle - T_k^2.
\end{align}
These higher order fluctuations cannot be expressed by the mean velocity and temperature in the case of non-Gaussian distributions. The equations governing the dynamics of ${\boldsymbol \theta}=(\theta_x,\theta_y)$ can be derived for nonlinear friction functions by considering the evolution of higher moments ($n>2$). 

For the mixed moments $\langle v_x^n v_y^m \rangle$ we obtain the following expressions:
\begin{subequations}\label{eq:covariance}
\begin{align}
\langle v_x \ v_y \rangle &= u_x \ u_y \\
\langle v_x \ v_y^2 \rangle &= u_x \ u_y^2 + u_x \ T_y\\
\langle v_x^2 \ v_y \rangle &= u_x^2 \ u_y + u_y \ T_x\\
\langle v_x^2 \ v_y^2 \rangle &= u_x^2 \ u_y^2 + u_x^2 \ T_y + u_y^2 \ T_x + T_x \ T_y. 
\end{align}
\end{subequations}
We insert Eqs. \eqref{eq:simplemoments} and \eqref{eq:covariance} in the equation for the moment dynamics \eqref{eq:RHva2d_mom} and, by carrying out the calculations up to the second order ($n=2$), we arrive at the following set of partial differential equations
\begin{subequations}
\begin{align}
 \frac{\partial}{\partial t} \rho  = &  -{\nabla}_{\bf r} \left(\rho  {\bf u}\right) \label{eq:continuity},\\
 \frac{\partial u_x}{\partial t} + {\bf u}  {\nabla}_{\bf r} u_x  =& \  \alpha  u_x - \beta  u_x  \left({\bf u}^2 + 3 T_x + T_y\right) +\mu(u_{\varepsilon,x}-u_x)\nonumber\\
 &\  - \frac{\partial T_x}{\partial x} - \frac{T_x}{\rho} \ \frac{\partial \rho}{\partial x}, \\
 \frac{1}{2} \left(\frac{\partial T_x}{\partial t} + {\bf u}\nabla_{\bf r} T_x\right) =&\  (\alpha-\mu)  T_x - \beta T_x \left({\bf u}^2 + 2 u_x^2 + T_x + T_y\right)  - \beta  \theta_x \nonumber \\
 &\  + D - T_x \frac{\partial u_x}{\partial x}  \ .
\end{align}
\label{eq:RHva2d_dgls}
\end{subequations}
Here, we have only given the equation for $x$-components of the mean velocity and temperature; the corresponding equations for the $y$-component can be directly obtained by interchanging $x$ and $y$. 


In the following we will focus on the spatially homogeneous system, which corresponds to the case of global coupling. In this case the above equations simplify to ordinary differential equations for the mean field velocity and temperature ($x$-component):
\begin{subequations}\label{eq:MF_RH2dfull}
\begin{align}
\frac{d u_x}{dt} = & \alpha  u_x - \beta  u_x  \left({\bf u}^2 + 3 T_x + T_y\right),  \\
\frac{d u_y}{dt} = & \alpha  u_y - \beta  u_y  \left({\bf u}^2 + T_x +  3 T_y\right), \\
\frac 12 \frac{d T_x}{dt} = &  (\alpha-\mu)  T_x - \beta T_x \left({\bf u}^2 + 2 u_x^2 + T_x + T_y\right)  - \beta  \theta_x + D,\\
\frac 12 \frac{d T_y}{dt} = &  (\alpha-\mu)  T_y - \beta T_y \left({\bf u}^2 + 2 u_y^2 + T_x + T_y\right)  - \beta  \theta_y + D.
\end{align}
\end{subequations}
We perform a closure of the moment equations by setting $\theta_{x,y}=0$ and obtain a self-consistent set of ODE's. This is a reasonable approximation at low noise intensities $D$. In general, finite higher order fluctuations $\theta_{x,y}>0$ have an inhibiting effect on the temperature ${\bf T}$, which for sufficiently large $D$ may be not negligible.  

We can further simplify the above set of equations by choosing a reference frame where $u_x=u_{||}=u$ corresponds to the mean field velocity, whereas the orthogonal component vanishes $u_y=u_{\bot}=0$:
\begin{subequations}\label{eq:MF_RH2dred}
\begin{align}
\frac{d u}{dt} = & \alpha  u - \beta  u  \left(u^2 + 3 T_\| + T_\bot\right)  \\
\frac 12 \frac{d T_\|}{dt} = &  (\alpha-\mu)  T_{\|} - \beta T_\| \left(3 u^2 + T_\| + T_\bot\right)   + D\\
\frac 12 \frac{d T_\bot}{dt} = &  (\alpha-\mu)  T_\bot - \beta T_\bot \left(u^2 + T_\| + T_\bot\right)   + D
\end{align}
\end{subequations}
with $T_\|$ and $T_\bot$
being the temperature components parallel and perpendicular to the mean field direction of motion.
In the stationary disordered state ($u=0$), the temperature components can be easily calculated from \eqref{eq:MF_RH2dred}
and the corresponding solution reads
\begin{subequations}\label{eq:MF_RH2d_dissol}
\begin{align}
u_{1}= & 0, \\
T_{\| ,1}=T_{\bot,1}= 
T_1 = & \frac{\alpha-\mu + \sqrt{(\alpha-\mu)^2 +  8 \beta D}}{4 \beta} . 
\end{align}
\end{subequations}
In the case of vanishing noise $D=0$, the ordered solution can be immediately obtained as $u=\sqrt{\alpha/\beta}$ and $T_\|=T_\bot=0$. For $D>0$, it is evident that the temperature component parallel to the direction of motion is smaller than the perpendicular one: $T_\|<T_\bot$. 

For the general ordered state, with $u>0$ and $D>0$, we were so far not able to obtain explicit stationary solution for $u$, $T_\|$ and $T_\bot$ of the above ODE system \eqref{eq:MF_RH2dred} but the stable and unstable solutions can be determined by a numerical continuation methods as, for example, provided by the numerical software XPPAUT/AUTO97 \citep{doedel_auto:_1981,ermentrout_simulating_2002}.  

\begin{figure}
\begin{center}
  \includegraphics[width=0.8\linewidth]{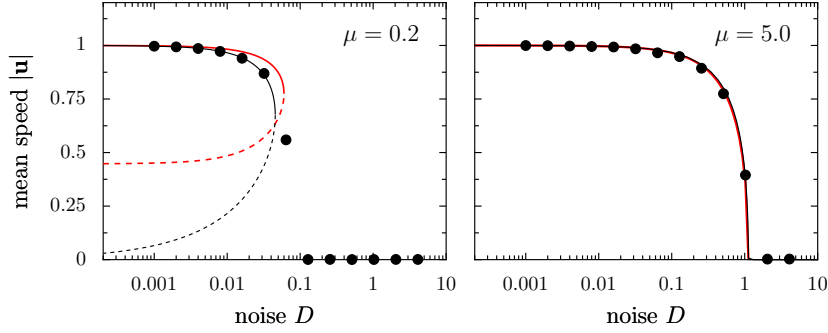} 
\end{center}
\caption{Comparison of the mean field speed $|u|$ obtained from Langevin simulations (symbols) of the RH-model in two spatial dimensions with the results of the mean field theory for the homogeneous case. The black lines represent the solutions obtained from the full system of mean field ODE's. The red lines represent the mean field solutions from the reduced system. The stable solutions are shown as solid lines, whereas dashed lines indicate the unstable solution. The simulations were performed with periodic boundary condition and with the disordered state as initial condition. Other parameters: $\alpha=1$, $\beta=1$, $L=200$, $\varepsilon=20$, $N=4096$.\label{fig:MF_RH2d_U}}
\end{figure}
\begin{figure}
\begin{center}
  \includegraphics[width=0.8\linewidth]{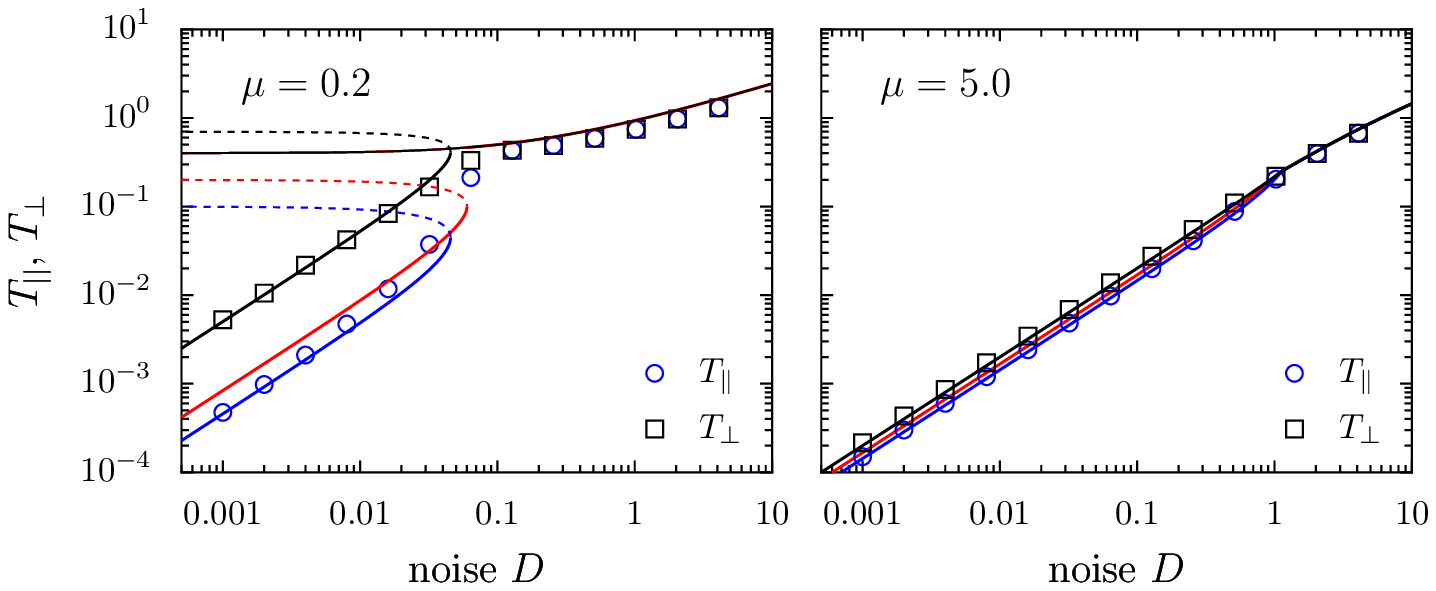} 
\end{center}
\caption{Numerical results for stationary temperature components, $T_\|$ (blue/dark gray circles) and $T_\bot$ (black squares), obtained from Langevin simulations of the RH-model with mean field predictions for weak velocity alignment (left) and strong velocity alignment (right). The solid (dashed) lines represent stable (unstable) solutions. The solutions of the full system for $T_\|$ ($T_\bot$) are indicated by blue/dark gray (black) lines. The red lines show the mean field solutions from the reduced system.   
The simulations were performed with periodic boundary condition and with the disordered state as initial condition. Other parameters: $\alpha=1$, $\beta=1$, $L=200$, $\varepsilon=20$, $N=4096$.\label{fig:MF_RH2d_T}}
\end{figure}

A possible ansatz to find an explicit solution is the reduction of the dimensions in the problem: We use the fact that at a fixed time $t$ we may always find a coordinate frame where $u_x=u_y=\tilde u$. In this coordinate frame due to the symmetry of the involved equations we obtain also $T_x=T_y=\tilde T$. With this ansatz we reduce the original system \eqref{eq:MF_RH2dfull} from a four dimensional system of ODE's to a two dimensional system in $\tilde u$ and $\tilde T$:
\begin{subequations}\label{eq:MF_RH2dreduced}
\begin{align}
\frac{d \tilde u}{dt} = & \alpha  \tilde u - \beta  \tilde u  \left(2 \tilde u^2 + 4 \tilde T\right)  \\
\frac 12 \frac{d \tilde T}{dt} = &  (\alpha-\mu)  \tilde T - \beta \tilde T \left(4 \tilde u^2 + 2 \tilde T\right) + D 
\end{align}
\end{subequations}
This gives us a system of equations similar to the one-dimensional case, where we can analytically determine the stationary solutions for $d\tilde u/dt=d\tilde T/dt=0$ to:
\begin{subequations}
\begin{align}
\tilde u_{1,2} & = 0, \\ 
\tilde T_{1,2} & = \frac{\alpha-\mu \pm \sqrt{(\alpha-\mu)^2 +  8 \beta D}}{4 \beta}, \\
\tilde u_{3,4} & = \pm \frac{\sqrt{2 \alpha -\mu + \sqrt{(\alpha+\mu)^2 - 24 \beta D}}}{\sqrt{6 \beta}}, \\
\tilde T_{3}=T_{4} & = \frac{\alpha + \mu - \sqrt{(\alpha+\mu)^2 - 24 \beta D}}{12 \beta}, \\
\tilde u_{5,6} & = \pm \frac{\sqrt{2 \alpha -\mu - \sqrt{(\alpha+\mu)^2 - 24 \beta D}}}{\sqrt{6 \beta}}, \\
\tilde T_{5}=T_{6} & = \frac{\alpha + \mu + \sqrt{(\alpha+\mu)^2 - 24 \beta D}}{12 \beta}. 
\end{align}
\end{subequations}
The type of the different solutions is completely analogous to the one-dimensional system: 
\begin{center}
\begin{tabular}{rl}
$\tilde u_{1,2}$, $\tilde T_1\ \ $ : & disordered solution \\
$\tilde u_{3,4}$, $\tilde T_{3,4}$ : & stable ordered solutions \\
$\tilde u_{5,6}$, $\tilde T_{5,6}$ : & unstable ordered solutions 
\end{tabular}
\end{center}
The structure of the bifurcation diagram of the reduced system \eqref{eq:MF_RH2dreduced} for $d=2$ is the same as for $d=1$ \citep{romanczuk_collective_2010} with different critical noise intensities determining the stability of the disordered and ordered solutions:
\begin{align*}
D^{(2d)}_{d,\rm crit} & = \frac{\alpha(2\mu -\alpha)}{8\beta}, \\
D^{(2d)}_{o,\rm crit} & = \frac{(\alpha+\mu)^2}{24\beta}.  
\end{align*}
The three different regimes obtained from the reduced system can be summarized as: 
\begin{enumerate}
\item weak alignment ($0 < \mu < \frac{\alpha}{2}$):   
\begin{enumerate}
\item bistability for weak fluctuations $D<D_{o,\rm crit}$ 
\item only disordered solution stable for strong fluctuations $D>D_{o,\rm crit}$
\end{enumerate}
\item intermediate alignment ($\frac{\alpha}{2}<\mu<2\alpha $):  
\begin{enumerate}
\item only ordered solution stable for weak fluctuations $D<D_{d,\rm crit}$
\item bistability for intermediate noise $D_{d,\rm crit}<D<D_{o,\rm crit}$
\item only disordered solution stable for strong fluctuations  $D>D_{o,\rm crit}$
\end{enumerate}
\item strong alignment ($\mu > 2\alpha$): 
\begin{enumerate}
\item only ordered solution stable for weak fluctuations $D<D_{o,\rm crit}$ 
\item only disordered solution stable for strong fluctuations $D>D_{d,\rm crit}$
\end{enumerate}
\end{enumerate}
%
The comparison of the stationary solutions of the reduced systems with the corresponding solutions of the full systems obtained with XPPAUT/AUTO reveals differences at low velocity-alignment strengths $\mu$ (Fig. \ref{fig:MF_RH2d_U},\ref{fig:MF_RH2d_T}). The velocity of the stable ordered solution of the full system decreases more strongly and exhibits an earlier breakdown with increasing $D$. Furthermore, from the position of the disordered branch, it can be deduced that the basin of attraction of the ordered state for the full system at low noise $D$ is larger than for the reduced two dimensional system. At large $\mu$ the differences between the two types of mean field solution vanish and the reduced system gives a good approximation as shown in Fig. \ref{fig:MF_RH2d_U}. 
The reason for the discrepancy between the two mean field solutions at low $\mu$ is the fact that the reduction of the system dimensions ``throws away'' all information about the asymmetry of temperature components parallel and perpendicular to the mean velocity. But at large $\mu$ both temperature coefficients are dominated by the $-\mu T_k$ term, so that the asymmetry in the temperature components becomes negligible as shown in Fig. \ref{fig:MF_RH2d_T}. Please note, that in the two-dimensional phase-space projections used in Fig. \ref{fig:MF_RH2d_T}, distinct non-overlapping solution with $|u|=0$ and $|u|>0$ appear on top of each other. 

In general, without knowing the temperatures $T_\|$ and $T_\bot$, the mean speed as the order parameter can be written as   
\begin{align}
|u| = \sqrt{v_0^2 - 3 T_\| - T_\bot}
\end{align} 
here we used $v_0^2=\alpha/\beta$.
In the limit of large $\mu$ close to the critical noise, where $\alpha,\beta \ll \mu, D$, we may approximate the temperature as  $T_\|=T\bot=T=D/\mu$ and we obtain a simple expression for the ordered state
\begin{align}\label{eq:MF_RH2d_ulimit}
|u| = \sqrt{v_0^2 - \frac{ 4D}{\mu} }.
\end{align}
In this limit, the critical noise can be approximated as $D_{d,\rm crit}\approx v_0^2\mu/4=D_\text{crit}$ and the above equation may be rewritten as:
\begin{align}\label{eq:MF_RH2d_ulimit_2}
|u| = 2\mu^{-\frac 12 }(D_\text{crit} - D)^{\frac 12},
\end{align}
which is the standard form of the order parameter for a continuous (second order) phase transition.

In order to check the analytical results, we have performed large scale Langevin simulation of the microscopic system with periodic boundary condition.
In two dimensions, the direction of motion in the ordered state ${\bf u}/|{\bf u}|$ can freely diffuse. Due to always present fluctuations in a finite system the direction of motion changes in time. Thus, the ordered state is characterized in simulations through the non-vanishing mean speed of the system:
\begin{align}
\langle | {\bf u}| \rangle =\left\langle \sqrt{\left(\frac{1}{N} \sum_{i=1}^{N} {\bf v}_i\right)^2} \right\rangle.
\end{align} 
Here, $N$ is the total number of individuals and $\langle \cdot \rangle$ indicates the temporal average after the system reached the stationary state.      

The results of microscopic Langevin simulations in two spatial dimensions $d=2$ at high densities and $L/\varepsilon=10$ are in a good agreement with the semi-analytical results for the full mean field system at low $\mu$ and with both solutions types at large $\mu$. The temperature components $T_\|$ and $T_\bot$ differ significantly at low $\mu$ and the solution of the reduced system is not able to describe the system correctly. But for large $\mu$ the differences between the different components become negligible and the different stationary mean field solution and the numerical results collapse practically on a single line (Figs. \ref{fig:MF_RH2d_U}, \ref{fig:MF_RH2d_T}). 

Even if the corresponding basin of attraction of the disordered state is reduced in comparison to the one-dimensional case \citep{romanczuk_collective_2010}, the mean field theory predicts nevertheless its stability at low $\mu$ and low noise $D$. But in contrast to the one-dimensional case,  we were not able to observe a stable stationary disordered state in numerical simulations even at very low $\mu$ and low but non-vanishing $D$ (Fig. \ref{fig:MF_RH2d_U}). This is a consequence of neglecting all higher order fluctuations $\theta_k$ in our mean field ansatz. In two dimensions it has a stronger impact on the stability of the disordered state than in the one-dimensional case in the respective parameter region. In two spatial dimensions at low $\mu$ and low $D$ there is no energetic barrier between different directions of motion. Here, the individuals can change their direction of motion by continuous angular drift or diffusion. Thus, any finite fluctuation in ${\bf u}$ at vanishing noise will be amplified and eventually will lead to perfect alignment (see also \citealp{ebeling_swarm_2008}).       

\begin{figure}
\begin{center}
  \includegraphics[angle=-90,width=0.49\linewidth]{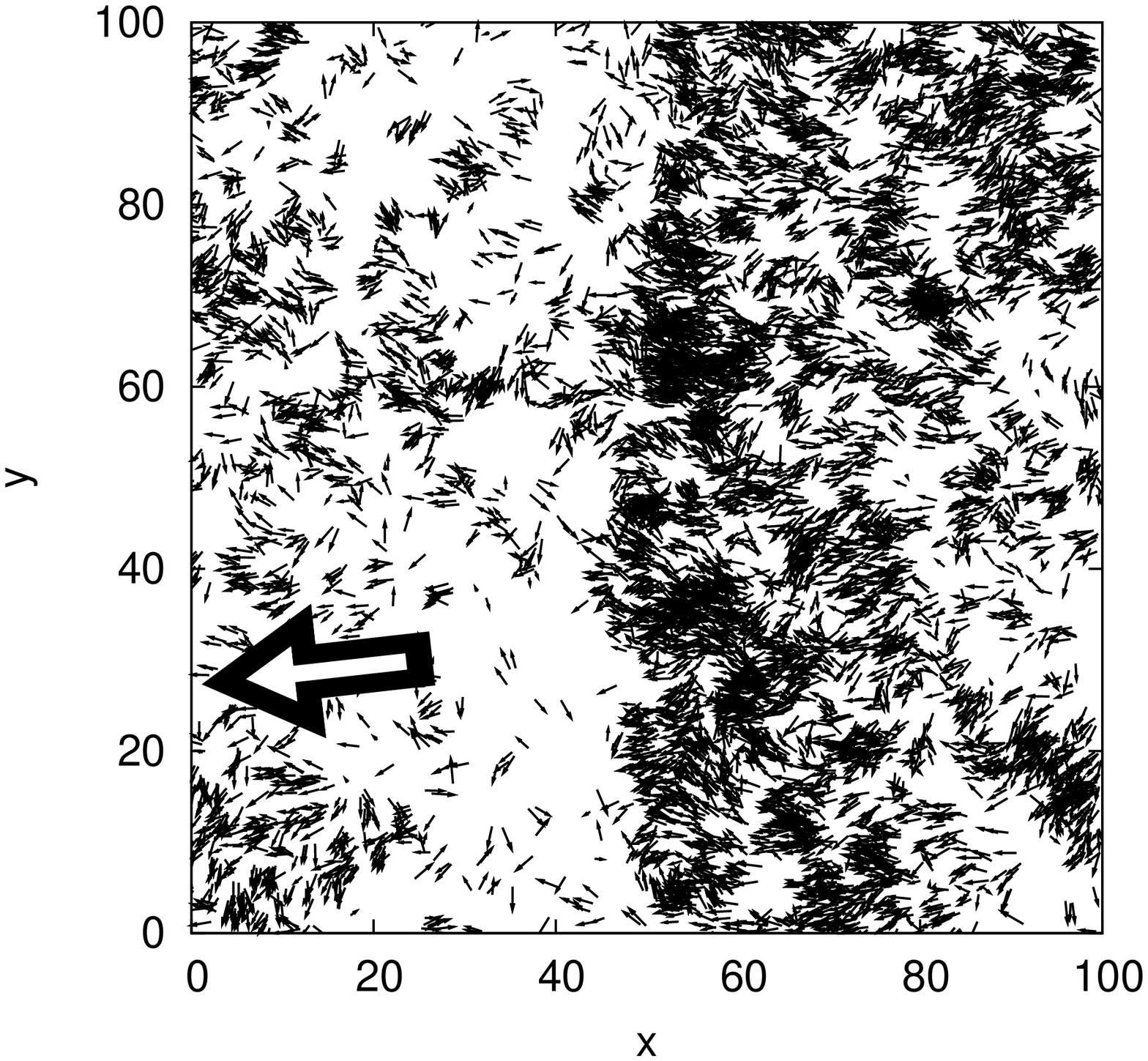} 
  \includegraphics[angle=-90,width=0.49\linewidth]{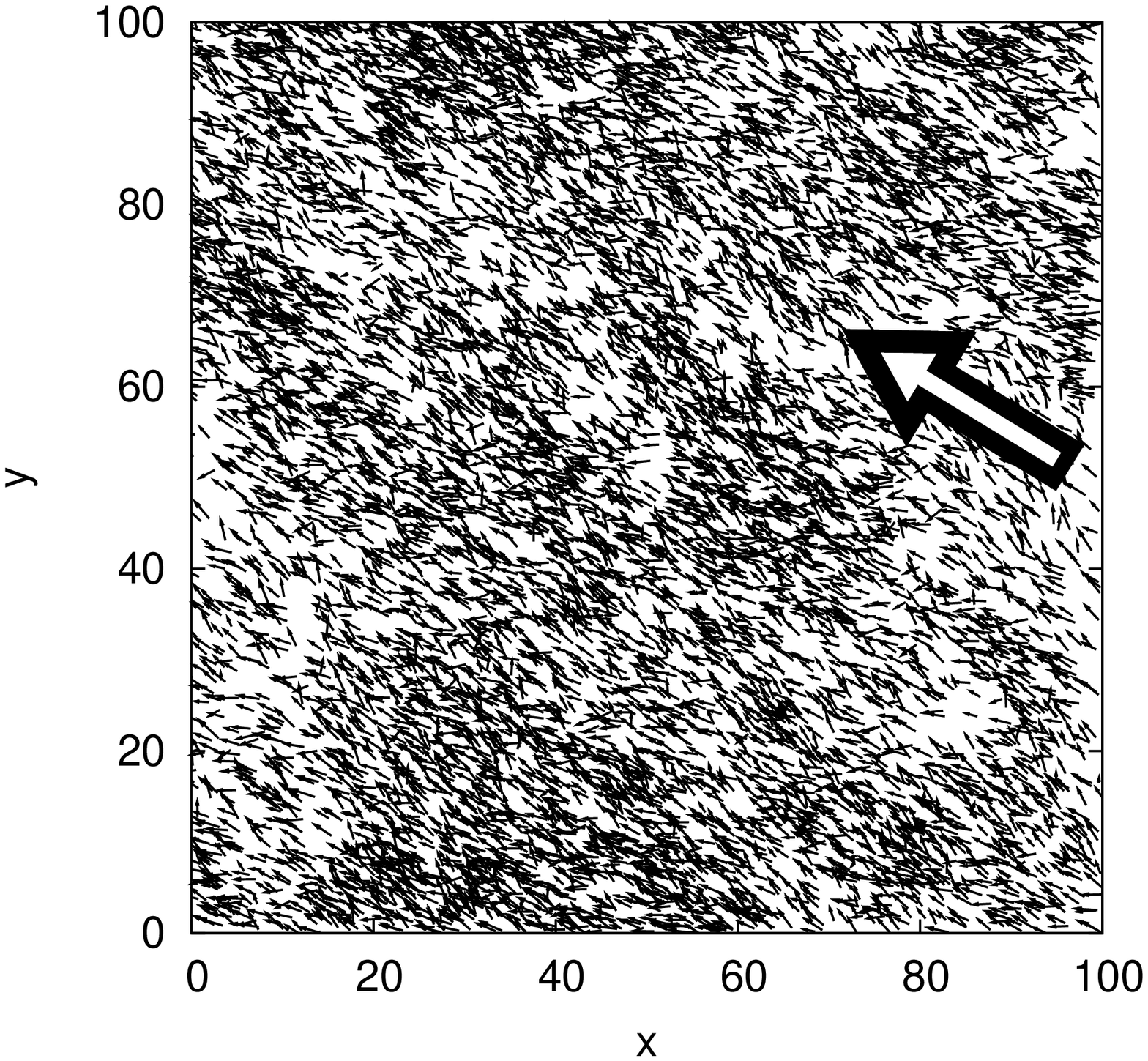} 
\end{center}
\caption{Snapshots of the ordered state of active Brownian particles with RH-friction and velocity alignment for two different values of $\varepsilon=1.0$ (left) and $5.0$ (right). The large arrow indicates the mean direction of motion. (Other parameter values: $N=8192$, $\mu=1.0$, $D=0.1$, $L=100$)\label{fig:snapshots_va}}
\end{figure}

Recently, it was shown that the spatially homogeneous state is unstable in systems of interacting self-propelled particles \citep{bertin_boltzmann_2006,bertin_microscopic_2009,simha_statistical_2002,simha_hydrodynamic_2002,ihle_kinetic_2011}. Thus, the assumption of a spatially homogeneity is only an approximation, which is justified only for sufficiently large $\varepsilon$ (upper panel, Fig. \ref{fig:snapshots_va}). For $L\gg\varepsilon$ strong density inhomogeneities appear, such as travelling bands (lower panel, Fig. \ref{fig:snapshots_va}), which affect the global behavior of the system. 

\section{Schienbein-Gruler friction function}
As a second example, we consider a system of active Brownian particles with linear speed dynamics studied by \citet{schienbein_langevin_1993}. Here we use a corresponding friction function as introduced by \citet{erdmann_brownian_2000} in two spatial dimensions:
\begin{align}\label{eq:SGva2d_frifun}
-\gamma({\bf v}){\bf v}=-\alpha \left(1-\frac{v_0}{|{\bf v}|}\right){\bf v},
\end{align}
where, $v_0$ is the preferred speed of individuals (stationary speed), and $\alpha$ is now the inverse relaxation time of the velocity dynamics.
Please note that the above friction/propulsion term, shows a discontinuity at the origin ${\bf v}={\bf 0}$, which may be eliminated by taking into account the possibility of backwards motion of individuals with respect to their heading \citep{romanczuk_active_2011} neglected by \citet{schienbein_langevin_1993}. But the much most important property of the above friction function, irrespective of the details at vanishing speed, is its linear behavior around the stationary speed $v_0$.

Following the same approach as in the previous examples we arrive at the following moment equation for the $x$-component:
\begin{align}\label{eq:SGva2d_mom}
\frac{\partial}{\partial t}(\rho v_x^{n}) = & 
			- \frac{\partial}{\partial x} \left(\rho \ \langle v_x^{n+1} \rangle\right) - \frac{\partial}{\partial y} \left(\rho \ \langle v_x^n \ v_y \rangle\right)  \nonumber \\
 	& + n \  \rho \left[ \alpha \left( \left\langle \frac{v_x^n}{\sqrt{v_x^2+v_y^2}} \right\rangle -    \langle v_x^{n} \rangle  
	\right)   
	+ \mu ( u_{\varepsilon,x} \langle v_x^{n-1} \rangle - \langle v_x^n \rangle ) \right] \nonumber \\
 & +n \ \left(n-1\right) \ D \ \rho \ \langle v_x^{n-2} \rangle.
\end{align}
The above equation differ from the one obtained in the Rayleigh-Helmholtz case  \eqref{eq:RHva2d_mom} only in terms associated with the friction function. Due to the absolute value of the velocity $|{\bf v}|$ in \eqref{eq:SGva2d_frifun}, we obtain in the moment evolution equation not only linear combination of moments but also the term $\langle v_x^n/\sqrt{v_x^2+v_y^2}\rangle$, which we have to take care of. Here, we make the following approximation:
\begin{align}\label{eq:sg_momappr}
\left\langle \frac{v_x^n}{\sqrt{v_x^2+v_y^2}} \right\rangle \approx \frac{\langle v_x^n\rangle}{\left\langle \sqrt{v_x^2+v_y^2}\right\rangle} \approx \frac{\langle v_x^n\rangle }{\sqrt{u_x^2+u_y^2+T_x+T_y}}.
\end{align}
This approximation, results in a closure of the final mean field equations.
For clarity, we will use the following abbreviation for the denominator in the above expression 
$$V_T:=V_T({\bf u},{\bf T})=\sqrt{u_x^2+u_y^2+T_x+T_y}.$$
Thus, we obtain the following system of mean field equations for the $k$th-component of the mean field velocity and temperature ($k=x,y$):
\begin{subequations}
\begin{align}\label{eq:SGva2d_dgls}
\frac{\partial u_k}{\partial t} + {\bf u}  {\nabla}_{\bf r} u_k  = &  \left(\frac{\alpha v_0}{V_T}-\alpha\right)u_k +\mu(u_{\varepsilon,k}-u_k)
 - \frac{\partial T_k}{\partial x} - \frac{T_k}{\rho} \ \frac{\partial \rho}{\partial x}, \\
\frac{1}{2} \left(\frac{\partial T_k}{\partial t} + {\bf u}\nabla_{\bf r} T_k\right) =& \left(\frac{\alpha v_0}{V_T}-\alpha-\mu\right)  T_k + D - T_k \frac{\partial u_k}{\partial x}, 
\end{align}
\end{subequations}
\noindent which together with the continuity equation \eqref{eq:continuity} determine the evolution of the mean field system. 

By considering again the simplest case of a spatially homogeneous system, we arrive at the following set of ordinary differential equations
($k=x,y$ or $\|,\bot$):
\begin{subequations}\label{eq:SGva2d_homog0}

\begin{align}
\frac{d u_k}{dt} = &  \left(\frac{\alpha v_0}{V_T}-\alpha\right)u_k, \\
\frac{1}{2}\frac{d T_k}{dt} = & \left(\frac{\alpha v_0}{V_T}-\alpha-\mu\right)  T_k + D. 
\end{align}
\end{subequations}
From the symmetry of the temperature equations it follows directly $T_x=T_y$ or by choosing the appropriate reference frame $T_\|=T_\bot$. Using the mean velocity squared  $u^2=u_x^2+u_y^2$ as a variable instead of the individual components ($u_x$, $u_y$) the above equations simplify to:
\begin{subequations}\label{eq:SGva2d_homog}
\begin{align}
\frac{d}{dt} u ^2 = &  2\alpha\left(\frac{v_0}{V_T}-1\right)u^2, \label{eq:MFSG_2d_msqrEq}\\
\frac{1}{2}\frac{d T_k}{dt} = & \left(\frac{\alpha v_0}{V_T}-\alpha-\mu\right)  T_k + D.
\end{align}
\end{subequations}
There are two stationary solution of Eq. \ref{eq:MFSG_2d_msqrEq} in terms of the temperature components reads:
\begin{subequations}
\begin{align}
u^2 & =0,  \\u^2& =v_0^2-T_\|-T_\bot \label{eq:MFSG_2d_msqrSol}.
\end{align}
\end{subequations}
The full system of equations has in total three stationary solution: a single ordered state solution with $|u|=\sqrt{u^2}>0$ and two disordered solutions with $|u|=0$. The single ordered solution reads:
\begin{align}\label{eq:MF_SG2d_uord}
 |u|_1   =  \sqrt{v_0^2-\frac{2D}{\mu}},\qquad  T_{k,1}  =  \frac{D}{\mu}, 
 \end{align}
 whereas two stationary disordered solutions are given as:
 \begin{align}
 |u|_{2,3} &  =  0, \\
 T_{k,2}  & =  \frac{D}{\alpha + \mu}+\left(\frac{\alpha v_0}{2(\alpha+\mu)}\right)^2 \left[ 1 + \sqrt{ 1+\frac{8 D (\alpha +\mu)}{\alpha^2v_0^2}}\right], \label{eq:MFSG_2d_Td} \\
 T_{k,3}  & =  \frac{D}{\alpha + \mu}+\left(\frac{\alpha v_0}{2(\alpha+\mu)}\right)^2 \left[ 1 - \sqrt{ 1+\frac{8 D (\alpha +\mu)}{\alpha^2v_0^2}}\right].  
\end{align}
We identify $T_{k,2}$ with the temperature of the disordered phase. The second disordered solution $T_{k,3}$ is not considered, as it is always less stable than $T_{k,2}$ and yields unphysical result of vanishing temperature of the disordered phase in the limit $D,\mu \to0$. For noninteracting self-propelled particles, for which only the disordered state exists, the temperature has to coincide with the average kinetic energy per particle: $v_0^2/2$. It can be easily shown that $T_{k,2}$ satisfies this condition in the respective limit. 


\begin{figure}
\begin{center}
  \includegraphics[width=0.9\linewidth]{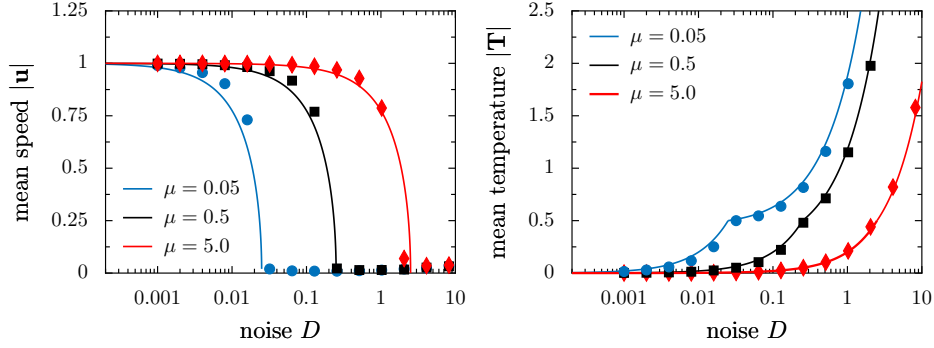} 
\end{center}
\caption{Comparison of the simulation and analytical results for the mean field speed $|u|$ (left) and temperature $|{\bf T}|$ (right) versus noise strength $D$ for the Schienbein-Gruler friction.  The results of Langevin simulations are shown as symbols. The solid lines represent the stationary solutions according to Eq. \eqref{fig:MFSG2d_fullsol}. The simulations were performed with periodic boundary condition and with the disordered state as initial condition. Simulation parameters: $\alpha=1.0$,$v_0=1.0$, $\varepsilon=10.0$, $L=100.0$, $N=16384$.
\label{fig:MF_SG2d_vs_D}}
\end{figure}
\begin{figure}
\begin{center}
  \includegraphics[width=0.9\linewidth]{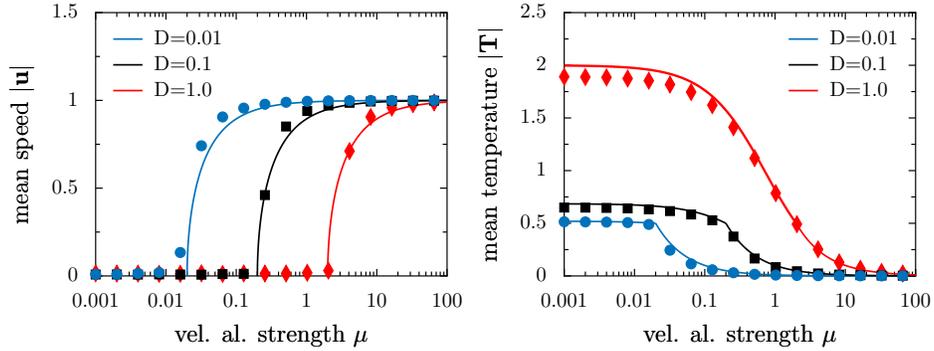} 
\end{center}
\caption{Comparison of the simulation and analytical results for the mean field speed $|{\bf u}|$ (left) and mean temperature $|{\bf T}|$ (right) versus velocity alignment $\mu$ for the Schienbein-Gruler friction. The results of Langevin simulations are shown as symbols. The solid lines represent the stationary solutions according to Eq. \eqref{fig:MFSG2d_fullsol}. The simulations were performed with periodic boundary condition and with the disordered state as initial condition. Simulation parameters: $\alpha=1.0$,$v_0=1.0$, $\varepsilon=10.0$, $L=100.0$, $N=16384$.}
\label{fig:MF_SG2d_vs_mu}
\end{figure}

For noise intensities below the critical noise intensity 
\begin{align}
D_{\rm crit}=v_0^2\mu/2,
\end{align} 
the homogeneous system is in the ordered state ($|u|=|u|_1$, $T_k=T_{k,1}$), whereas above the critical noise intensity no collective motion takes place ($|u|=|u|_2$, $T_k=T_{k,2}$). For vanishing noise, $D\to 0$, the temperature components $T_k$ vanish and the mean field speed is given as $|u|=v_0$. For $v_0=0$, the ordered state is always unstable and the temperature of the disordered state reduces to $T_k=D/(\alpha+\mu)$.
Inserting the critical noise intensity into \eqref{eq:MF_SG2d_uord} yields for the mean speed
\begin{align}
|u|=2^{\frac 12} \mu^{-\frac 12}(D-Dc)^{\frac 12}.
\end{align}
This corresponds to a second-order phase transition, not only in the limit of large $\mu$, as in the case of Rayleigh-Helmholtz friction.

The full stationary solution of the homogeneous system may be written as:
\begin{subequations}
\begin{align}\label{fig:MFSG2d_fullsol}
|u|=
\begin{cases} 
|u|_1 & \text{for} \quad D<D_\text{crit} \\
0 & \text{for} \quad D>D_\text{crit} \\
\end{cases} \\
T_k=
\begin{cases} 
T_{k,1} & \text{for} \quad D<D_\text{crit} \\
T_{k,2} & \text{for} \quad D>D_\text{crit} \\
\end{cases}
\end{align}
\end{subequations}
As $T_{k,1}=T_{k,2}=v_0^2/2$ for $D=D_\text{crit}$ the temperature is continuous but in general not continuously differentiable with respect to the bifurcation parameter (e.g. $D$, $\mu$).

\begin{figure}
\begin{center}
  \includegraphics[width=0.9\linewidth]{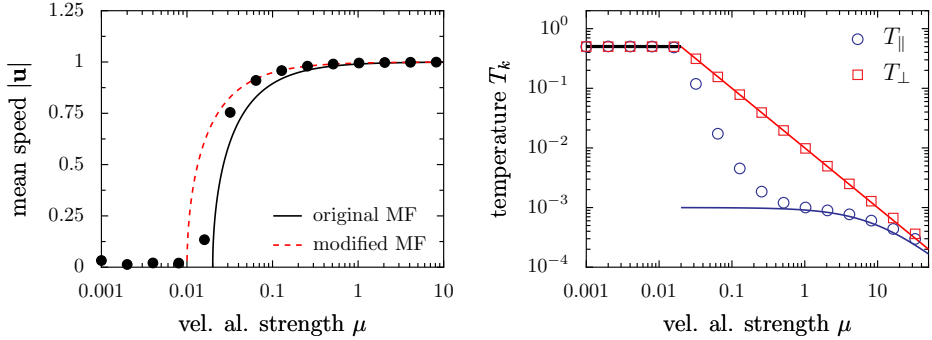} 
\end{center}
\caption{{\bf Left:} Comparison between $|\hat u|$ from Eq. \eqref{eq:MFSG_2d_modspeed}, $|u|$ from Eq. \eqref{eq:MF_SG2d_uord} and numerical results for the Schienbein-Gruler friction; {\bf Right:} The corresponding temperature components $\hat T_\|=D/(\alpha+\mu)$ (blue/dark gray line),  $\hat T_\bot=D/\mu$ (red/light gray line) and $T_{k,2}$ from \eqref{eq:MFSG_2d_Td}  (black line) together with results of numerical simulations (symbols). The Langevin simulations were performed with periodic boundary condition and global coupling. Simulation parameters: $\alpha=10.0$,$v_0=1.0$, $N=16384$.}
\label{fig:MF_SG2d_vs_mu_a10_global}
\end{figure}

In order to test the theoretical predictions, we have performed large scale Langevin simulation of the microscopic system either with global coupling $\varepsilon>L$, or with local coupling $\varepsilon=1/10 L$. At high densities,  $\tilde \rho =(N\varepsilon^2)/L^2\gg1$, both situations yield similar results due to the large correlations length in the ordered phase with $l_\text{corr}\gg\varepsilon$. Thus, also for local coupling the system can be assumed as spatially homogeneous, if the interaction range $\varepsilon$ is not to small with respect to the system size $L$. All simulations were performed with periodic boundary conditions and with random initial positions and velocities (disordered state). 
      
As shown in Figs. \ref{fig:MF_SG2d_vs_D} and  \ref{fig:MF_SG2d_vs_mu}, the numerical results for the mean speed and the total temperature 
$$|{\bf T}|=\sqrt{T_\|^2+T_\bot^2}=\sqrt{2}T_k$$ 
are in good agreement with the theoretical prediction. But a closer look reveals some systematic deviations. The theoretical results for the temperature $T$ are larger than the ones obtained in the simulations in particular at low $\mu$. The effect appears for both ordered and disordered state as well as for global and local coupling. This systematic difference vanishes for $v_0=0$, thus it may be associated with the moment approximation in the corresponding term of the friction function (Eq. \eqref{eq:sg_momappr}). Furthermore, as in the case of Rayleigh-Helmholtz friction, the temperature components parallel and perpendicular to the direction of motion are not equal, in contrast to the predictions of the mean field theory. Whereas the perpendicular component agrees well with the theoretical prediction of $T_\bot=D/\mu$, we observe in general smaller values of the parallel component.

Starting from the perfectly ordered state in the homogeneous state, it can be shown that for small noise intensities the components of the temperature can be approximated as:
\begin{align}\label{eq:MFSG_2d_modtemp}
\hat T_\|=\frac{D}{\alpha+\mu}, \quad \hat T_\bot=\frac{D}{\mu}.
\end{align} 
Using Eq. \ref{eq:MFSG_2d_msqrSol}, we obtain then an alternative solution for the mean field speed:
\begin{align}\label{eq:MFSG_2d_modspeed}
|\hat u| = \sqrt{v_0^2-D\left(\frac{1}{\alpha+\mu}+\frac{1}{\mu}\right)}=\sqrt{v_0^2-D\left(\frac{\alpha+2\mu}{\mu(\alpha+\mu)}\right)}
\end{align}
In the limit of small $\alpha$ or large $\mu$, this result converges towards the original mean field result. This solution ($|\hat u|$, $\hat T_\|$, $\hat T_\bot$) for the ordered state yields a better agreement with numerical results not to close to the critical point but deviates from the numerical solutions close to the onset of collective motion (see Fig. \ref{fig:MF_SG2d_vs_mu_a10_global}).    

\section{Conclusion and Outlook} 
We have shown how starting from the microscopic dynamics of active Brownian particles with velocity alignment a corresponding mean field theory for the coarse grained density $\rho({\bf r},t)$, velocity ${\bf u}({\bf r},t)$ and temperature ${\bf T}({\bf r},t)$ can systematically be derived. 

In general, depending on the details of the individual dynamics, approximations have to be applied in order to obtain a self-consistent set of equations. Here, we have discussed two different friction/propulsion functions governing the motion of individuals. We have compared the results of the mean field theory obtained under the assumption of a spatially homogeneous case, and have shown that they are in good agreement with numerics for a wide range of parameters. On the other hand, the approximations made in the derivation of the mean field theory, may result for some cases in deviations between the analytical predictions and numerics, which can be understood from a detailed analysis. 

The mean field approach provides important insights into collective dynamics due to velocity alignment. In the case of a spatially homogeneous system (e.g. global coupling), where the local velocity field the individuals are sensing, equals to the mean field velocity ${\bf u_\varepsilon}={\bf u}$ the alignment force term vanishes in the velocity equation. 
Thus, in this limiting case the velocity alignment acts only on the effective temperature of the active Brownian particle gas. It suppresses the individuals fluctuations around the current mean velocity leading to the stabilization of the ordered state with finite mean field velocity \citep{ebeling_swarm_2008,romanczuk_collective_2010}. 

Recently, \citet{yates_inherent_2009} analyzed the mean velocity and its fluctuation in the collective motion of locusts. Based on the observed dependence of mean velocity fluctuations they suggested that this implies  state dependent fluctuations (multiplicative noise) on the level of individuals. In our case the ``temperature'' $|{\bf T}|$ is a measure of fluctuations of individual velocities around the mean. Although it does not correspond directly to the fluctuation observable studied by Yates {\em et al.} in a finite system it is nevertheless related. The stationary values show a strong dependence of $|{\bf T}|$ on the corresponding mean velocity -- increasing $|{\bf T}|$ with decreasing$|{\bf u}|$ -- even in the absence of any multiplicative noise in the dynamics of individuals. This is a consequence of the nonlinear Fokker-Planck equation governing the dynamics of the probability distribution and the resulting effective state-dependent velocity potential. Thus our results suggest that in biological systems extreme caution should be taken when deducing individual behavior of interacting agents based on mean field measurements, and that there might exist other alternative explanation of the observed mean field dynamics of locusts than the suggested ``inherent noise'' \citep{yates_inherent_2009}.     

Furthermore, we have shown that for the homogeneous case in the limit of large coupling $\mu$, where the impact of the individual speed dynamics (friction function) in the temperature equations becomes negligible, the scaling of the mean speed as order parameter close to the critical noise reads: 
\begin{align}  
|u|\sim \left(\frac{D_\text{crit}-D}{\mu}\right)^{\frac 12}. 
\end{align}
Thus, in this limit the onset of collective motion in a spatially homogeneous system takes place via a continuous (second-order) phase transition, irrespective on the details of the velocity dynamics. This result agrees also with previous results obtained for a system of self-propelled particles with constant speed \citep{peruani_mean-field_2008}. 

In one dimension it was shown, that for weak coupling (low $\mu$) and a nonlinear Reyleigh-Helmholtz friction function, bistability of the ordered and disordered state is possible \citep{mikhailov_noise-induced_1999,romanczuk_collective_2010}. The corresponding mean field theory predicts a discontinuous transition from an initially ordered state of collective motion to the disordered state with increasing noise intensity for nonlinear friction functions. But the results of direct large-scale numerical simulations of the individual based model reveal deviations from the predicted behavior. 

A possible explanation could be an extremely small basin of attraction of the disordered solution in the respective parameter region and the always present finite fluctuations in Langevin simulations (largest system studied with global coupling: $N=32768$). But a more probable cause, is the approximation of the non-Gaussian probability distribution by a finite number of moments, and the corresponding unresolved impact of higher order fluctuations on the stability and bifurcation behavior of the system, which represents an interesting theoretical challenge for the future. Please note that a Gaussian approximation with $\theta_k=2T_k$ offers an alternative closure of moment equations for nonlinear friction function, but is unsuitable for the analysis of the disordered state at vanishing noise $D$ as it would imply also a vanishing temperature, which for active particles is certainly not the case.     
   
Our kinetic equations can also be used to analyze the stability of the spatially homogeneous state with respect to spatial perturbations. A corresponding instability in large scale collective motion has been predicted from the analysis of hydrodynamic equations of self-propelled particle systems \citep{bertin_boltzmann_2006,bertin_microscopic_2009,simha_statistical_2002,simha_hydrodynamic_2002,ihle_kinetic_2011} and large density fluctuations were also reported in numerical studies \citep{gregoire_onset_2004,ginelli_large-scale_2010,peruani_cluster_2010}. Similar density inhomogeneities can be observed in numerical simulations of our model for $\varepsilon \ll L$  as shown in Fig. \ref{fig:snapshots_va}. Thus, we should emphasize that our results for the homogeneous case hold strictly speaking only in the limit of global coupling and only as an approximation for local coupling, where the interaction range is not to small $\varepsilon$ in comparison to the system size $L$.

The approach presented in this paper, can be applied also to other swarming models, such as, the ``escape \& pursuit'' model introduced recently by the authors \citep{romanczuk_collective_2009}. The formulation of kinetic equations for interacting agents allow to address the question of how swarming affects population dynamics, in particular the population dispersal, at ecological length scales. A Velocity-alignment interaction between individuals may result in long range ordered collective motion, which corresponds to large scale population fluxes, and therefore the population dynamics can not be described by simple diffusion equations.   
  
We believe that the general approach of formulating individual based models in terms of stochastic differential equations and the derivation of macroscopic equations based on the corresponding nonlinear Fokker-Planck equation can be very useful to address many different problems of interacting agents in ecology. In particular, in the context of ecological modelling, it allows to establish a direct link between the individual based description and the macroscopic equation for systems of interacting agents. This link between the different levels of description is essential to gain a profound understanding of the self-organization on the population level but its direct mathematical derivation represents often a major challenge.

\end{document}